\documentclass[runningheads,fleqn]{cl2emult}

\usepackage{makeidx}  
\usepackage{graphicx} 
\usepackage{subeqnar} 
\usepackage{multicol} 
\usepackage{cropmark} 
\usepackage{phys}     
\makeindex            

\clearpage
\addcontentsline{toc}{section}{Index}
\printindex

\begin{document}
\title*{Molecular Dynamic Simulation of Directional Crystal Growth}
\toctitle{Molecular Dynamic Simulation of Directional Crystal Growth}
%
%
\titlerunning{Molecular Dynamic Simulation of Directional Crystal Growth}
%
\author{B.V.Costa\inst{1}
\and P.Z.Coura\inst{2}
\and O.N.Mesquita\inst{1}}
\authorrunning{B.V.Costa et al.}
%
%
\institute{Universidade Federal de Minas Gerais, 
Departamento de F{\'\i}sica
ICEx, C.P. 702, CEP 30123-970, Belo Horizonte - MG, Brazil
\and Universidade Federal de Juiz de Fora, 
Departamento de F{\'\i}sica}
\maketitle              

\begin{abstract}
We use molecular dynamic to simulate the directional growth of binary
mixtures. our results compare very well with analitical and experimental
results. This opens up the possibility to probe growth situations which
are difficult to reach experimentally, being an important tool for further
experimental and theoretical developments in the area of crystal growth.
\end{abstract}

\section{Introduction}

The rapid expansion of the use of high quality crystaline materials in optical
and eletronic devices has strongly stimulated research, both theoretical and
experimental, on dynamics of crystalization. Computer simulation has played an
important role on the development and understanding of crystal growth. During growth
the solid-fluid interface can display several interesting phenomena like segregation,
dynamical instabilities and pattern formation \cite{cross}. \\

A crystal can grow from the adjacent fluid by different mechanisms, depending on the
structure of the interface (rough or smooth), material purity, growth rates,
temperature gradients and so on. For a crystal to grow atoms or molecules must be
transported from the fluid towards the interface with a non-zero sticking probability.
besides that, the latent heat generated at the interface as well as the solute excess
segregated must be carried away. These requirements can be met in a controlled way by
puting the sample in an appropriated furnace submmited to a temperature gradient and
pulling it with constant speed toward the colder region. results of such experiments
have been compared with results of two dimensional models. Nijmeijr and
Landau\cite{marco} reported molecular dynamic (MD) simulation on laser heated pedestal
growth of fibers. Previous simulations consisted of kinetic models using Monte Carlo
techniques and numerical solution of transport equations\cite{weeks}. \\

\section{Simulation}

In this work we use MD to simulate the solidification of a two component system of a
solvent (particles of type a) and solute (particles of type b) interating via a
Lennard-Jones (LJ) potential. By tunning the LJ parameters we set the struture of the
interface (rough or smooth) and the segregation coefficient. We also simulate
morphological instabilities, the planar inteface becoming cellular and eventually
dendritic (Mullins-sekerka instability.). The simulation is carried out with all
particles interacting  through the LJ potential

\begin{equation}
\phi_{i,j}(r_{i,j}) = \epsilon_{i,j}
\left [
(\frac{\sigma_{i,j}}{r_{i,j}})^{12} -
(\frac{\sigma_{i,j}}{r_{i,j}})^{6} .
\right ]
\end{equation}

\begin{figure}
\includegraphics[width=.9\textwidth]{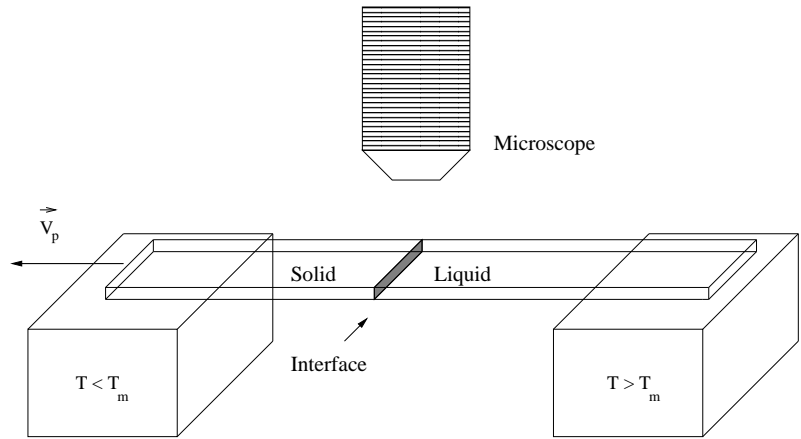}
\caption[]{Basic experimental set up for directional growth. The interface motion is
visualized using an optical microscope. $T_m$ is the melting temperature of the
mixture and $V_p$ is the pulling velocity.}
\label{eps1}
\end{figure}

The indexes $i$ and $j$ stands for particles in the positions $r_i$ and $r_j$
respectively and $r_{i,j} = |r_i - r_j|$. There are three types of interactions,
solvent-solvent, solvent-solute and solute-solute. In each case the LJ parameters are
labeled as $(a,a)$, $(a,b)$ and $(b,b)$. distance $r$, time $t$ and temperature $T$
are measured in units of $\sigma_{a,a}$,
$\sigma_{a,a}(m_a/\epsilon_{a,a})^{\frac{1}{2}}$ and $k_{B}/\epsilon_{a,a}$. initially
we distribute $N=n_x \times n_z$ particles over the two dimensional volume
$L_x \times L_z$. We assume periodic boundary conditions in the $x$ direction. In the
$z$ direction we divide the system in to two distinct regions, a solid and a fluid
one. In the solid region particles stand initially in their equilibrium position in a
total of $n_x \times n_{0z}$ particles. on the fluid region we distribute the rest of
the particles with the density chosen to be $\rho = 0.5\sigma_{a,a}^{-2}$, randomly
distributed in a triangular lattice and slightly dislocated from their equilibrium
position. We impose a temperature gradient along the $z$ direction using a velocity
renormalization approach. The system is divided in two regions, one cold which is set
to $T=0$ and the other to $T=T_h$ higher than the melting temperature $T_m$ which is
$T_m=0.4$ in our units. we let the system evolve in time for $N_e$ time steps of size
$\delta t$ until it reachs equilibrium. Once the equilibrium is reached we start
pulling the system in the $+z$ direction, at a pulling velocity $V_p$. The $-z_{max}$
layer works as a particle source, maintaining a constant flow to the material. The
basic experimental set up is shown in Fig.~\ref{eps1}

\begin{figure}
\includegraphics[width=.9\textwidth]{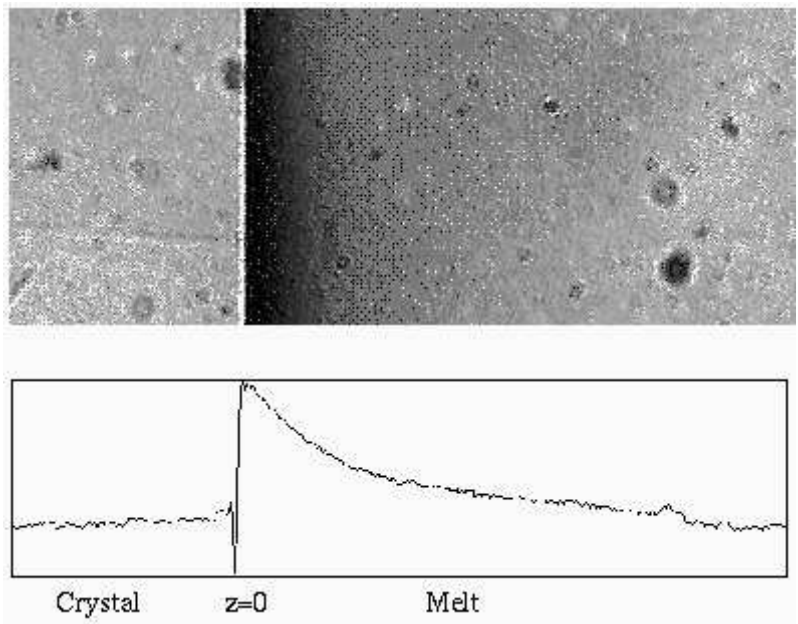}
\caption[]{Directional solidification of a binary mixture showing the solute
segregation at the interface. The experimental result is for the mixture caprolactane
(solvent) and methyl-blue (solute).}
\label{eps2}
\end{figure}

\begin{figure}
\includegraphics[width=.9\textwidth]{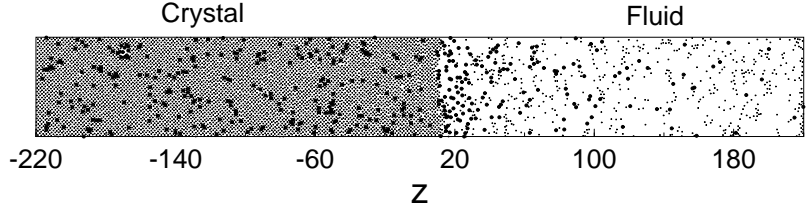}
\caption[]{Our simulation results using the Lennard-Jones potential describes in
detail the experiment.}
\label{eps3}
\end{figure}

In Fig.~\ref{eps2} we show an example of solute segregation during directional growth
o fthe binary mixture: caprolactane as solvent and methyl-blue as solute. In the top
part of Fig.~\ref{eps2} is shown an image of the crystal (left side) and melt (right
side), with maximum concentration of methyl-blue at the melt side of the interface.
From the  gray level of the image we obtain the methyl-blue concentration profile
across the sample, which in the melt, decays exponetially as a function of the
distance from the interface (bottom part of Fig.~\ref{eps2}). Fig.~\ref{eps3} shows
the simulation results done in a system initially with $27 \times 30$ particles,
solute concentration of $5\%$, pulling velocity $V_p = 0.004$, and LJ parameters given
by : $\epsilon_{a,b}=0.5$, $\epsilon_{b,b}=0.1$, $\sigma_{a,b}=1.0$, $\sigma_{b,b}=1.0$
and $m_b = 1.0$.

After averaging over many runs to improve statistics one obtains the steady-state
solute concentration profile represented as data points with error bars.

It is shown in Fig.~\ref{eps3} a theoretical result obtained from the diffusion
equation for the solute concentration in the fluid phase\cite{coura}

\begin{equation}
\frac{\partial c_f}{\partial t} =
D\frac{\partial^2 c_f}{\partial z^2} + V_f \frac{\partial c_f}{\partial z}
\end{equation}

where $D$ is the solute diffusion coefficient in the fluid phase and $V_f$ is the
velocity of the system of reference. The diffusion equation is supplemented with the
boundary conditions:

\begin{subeqnarray}
c_f & = & c_0 ~~~~~~~~~~~~~~~~ at ~~ z \rightarrow \infty \ts , \label{f1a}\\
(1-K)V_fc_f & = & -D\frac{d c_f}{d z} ~~~~~~~~ at ~~ z = 0  \ts .\label{f1b}
\end{subeqnarray}

The second condition is just the mass conservation at the interface and $K$ is the
ratio between the solute and solvent concentrations.

\section{Cellular Instability}

\begin{figure}
\includegraphics[width=.9\textwidth]{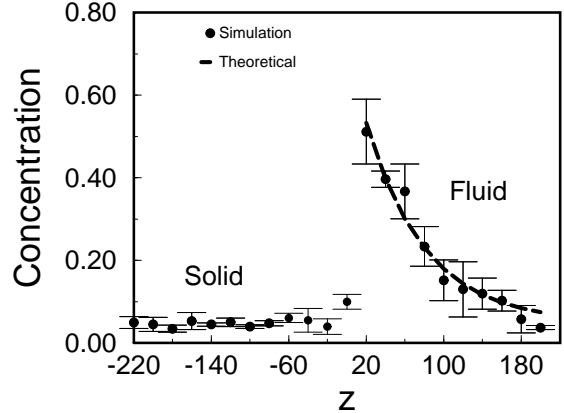}
\caption[]{The figure shows the concentration profile found in our simulation. It
is to be compared to Fig.~\ref{eps2} }
\label{eps4}
\end{figure}

With the LJ parameters chosen above the morphological instabilities were inhibited
during the cellular growth. We can stimulate instabilities by tunning the LJ
parameters, particularly by slightly varying the equilibrium position between
particles $a$ and $b$, i.e., changing $\sigma_{a,b}$. Fig.~\ref{eps4} show the effect
of varying this parameter.

\begin{figure}
\includegraphics[width=.8\textwidth]{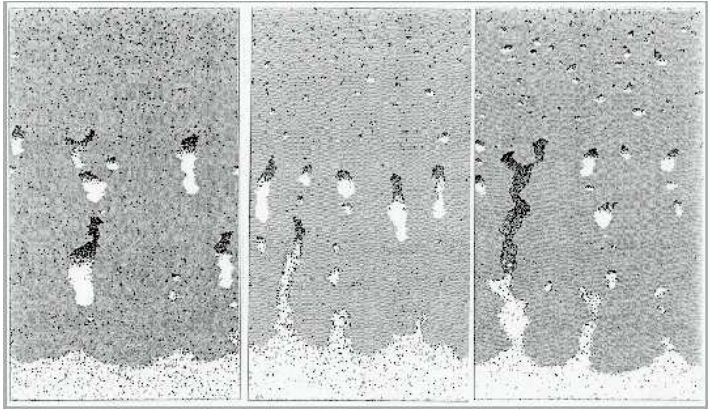}
\caption[]{Cellular instability originated by varying
           $\sigma_{a,b}$. From left to right 
           $\sigma_{a,b}=1.05,1.10$ and $1.20.$}
\label{eps5}
\end{figure}

For $\sigma_{a,b} = 1$ the interface is rough, becoming cellular and dendritic for
$\sigma_{a,b} > 1$. As shouls be expected a lower $\sigma_{a,b}$ sets the interface
smoother than larger values indicating that the system is very sensitive to geometric
factores. (We have also simulated some experiments fixing the value of $\sigma_{a,b}$
and varying $\epsilon_{a,b}$ and $\epsilon_{b,b}$, the results did not show any
qualitative change in their behavior. Raising $\sigma_{a,b}$ breaks the hexagonal
structure o fthe solid. The energy involved in such distortion is so large that the
interface segragate solute particles, bonds $(a,a)$ and $(b,b)$ are preferible than
$(a,b)$. Once particles of the type $b$ concentrate at the interface its melting point
is lowered and channels of $b$ particles are formed inside the crystal. Fig.~\ref{eps5}
Shows this effect.

We are now studying the stability \cite{figueiredo} of those interfaces. We start with
a smoth  interface then, we introduce a perturbation at the interface. It has been
shown that near the bifurcation from planar to cellular the time evolution of the most
instable Fourier mode of the perturbation can be described by a third order
Landau-amplitude equation. This however is the subject of a new research and will be
soon published elsewhere.

\section{Acknowledgements}
This work was partially supported by CNPq, FAPEMIG and FINEP. Numerical work was done
at the CENAPAD-MG/CO and in the LINUX parallel cluster at the {\sl Laborat\'orio de
Simula\c{c}\~ao} Departamento de F\'{\i}sica - UFMG.
%

\clearpage
\addcontentsline{toc}{section}{Index}
\flushbottom
\printindex


\begin{thebibliography}{7}
%
\addcontentsline{toc}{section}{References}


\bibitem{cross}M.C.Cross and P.Hohenberg, rev. mod. phys. {\bf 65}, 851(1993).
\bibitem{marco}M.J.P.Nijmeijer and D.P.Landau, Comput. Mater. Sci. {\bf 1}, 389(1993).
\bibitem{weeks}J.D.Weeks and G.H.Gilmer, Adv. Chem. Phys. {bf 40}, 157(1979),
D.P.Woodruff, The solid-liquid interface (Cambridge University Press, london, 1973).
\bibitem{coura} P.Z.Coura, O.N.Mesquita and B.V.Costa, Phys. rev B {bf 59}, 8345(1999).
\bibitem{figueiredo}J.M.A.Figueiredo, M.B.L.Santos, L.O.Ladeira and O.N.Mesquita,
Phys. Rev. Lett. {bf 71}, 4397(1993).

\end{thebibliography}
\end{document}